\documentclass[preprint,showpacs,superscriptaddress,floatfix]{revtex4}
\usepackage{amssymb}
\usepackage{graphicx}

\usepackage[centertags]{amsmath}
\usepackage{amsfonts,amssymb}
\usepackage{times}
\usepackage{subfigure}
\usepackage{verbatim}
\usepackage{color}

\def\eps{\varepsilon}

\def\RR{\mathbb{R}}  
\def\EE{\mathbb{E}}\def\PP{\mathbb{P}}

\def\<{\langle} \def\>{\rangle}

\newcommand{\barint}{\kern4pt \raise3.4pt\hbox{\vrule height.6pt
    width7pt} \kern-11pt \int}

\begin{document}

\title{Large Deviations in Fast-Slow Systems}

\author{Freddy Bouchet}
\affiliation{Laboratoire de Physique, ENS de Lyon, 46, all\'ee d'Italie,
  F-69364 Lyon cedex 07, FRANCE}
\author{Tobias Grafke}
\affiliation{Courant Institute of Mathematical Sciences, New York
  University, 251 Mercer Street, New York, NY 10012}
\author{Tom\'as Tangarife}
\affiliation{Laboratoire de Physique, ENS de Lyon, 46, all\'ee d'Italie,
  F-69364 Lyon cedex 07, FRANCE}
\author{Eric Vanden-Eijnden}
\affiliation{Courant Institute of Mathematical Sciences, New York
  University, 251 Mercer Street, New York, NY 10012}

\date{\today}

\begin{abstract}  
  The incidence of rare events in fast-slow systems is investigated
  via analysis of the large deviation principle (LDP) that
  characterizes the likelihood and pathway of large fluctuations of
  the slow variables away from their mean behavior -- such
  fluctuations are rare on short time-scales but become ubiquitous
  eventually. This LDP involves an Hamilton-Jacobi equation whose
  Hamiltonian is related to the leading eigenvalue of the generator of
  the fast process, and is typically non-quadratic in the momenta --
  in other words, the LDP for the slow variables in fast-slow systems
  is different in general from that of any stochastic differential
  equation (SDE) one would write for the slow variables alone.  It is
  shown here that the eigenvalue problem for the Hamiltonian can be
  reduced to a simpler algebraic equation for this Hamiltonian for a
  specific class of systems in which the fast variables satisfy a
  linear equation whose coefficients depend nonlinearly on the slow
  variables, and the fast variables enter quadratically the equation
  for the slow variables. These results are illustrated via examples,
  inspired by kinetic theories of turbulent flows and plasma, in which
  the quasipotential characterizing the long time behavior of the
  system is calculated and shown again to be different from that of an
  SDE.
\end{abstract}

\maketitle

\section{Introduction}
\label{sec:intro}

The evolution of many dynamical systems of interest involve the
interplay of fast and slow variables. Examples include planetary
motion, geophysical flows, climate-weather interaction models,
macromolecules, etc. In such systems one is typically interested in
the behavior of the slow variables on time-scales that are much longer
than that over which the fast variables evolve.  Under suitable
conditions, the fast variables are adiabatically slaved to the slow
ones, and the latter only feel the average effects of the former. When
this is the case the evolution of the slow variables on their natural
time-scale can be captured by a closed limiting equation for these
variables alone that is obtained by averaging out the effect of the
fast variables on the slow motions. This equation is valid in the
limit when the scale separation between the fast and slow variables is
infinitely wide, and it is an instance of the Law of Large Numbers
(LLN) in the present context.  Of course the scale separation is never
infinite in reality and the slow variables also experience
fluctuations above their average motion. Small fluctuations are
captured by the Central Limit Theorem (CLT) which provides a linear
stochastic differential equation (SDE) for the difference between the
slow variables and their mean. Large deviations from the LLN away from
the CLT scaling, on the other hand, can be characterized via a large
deviation principle (LDP). On time scales that are of order one with
respect to the clock of the slow variables, these large fluctuations
are rare events. However, the LDP also captures the long time behavior
of these variables when the effect of fluctuations is no longer
negligible and large deviations from the LLN are no longer rare.  This
is the case, for example, if the limiting equation given by the LLN
possesses multiple stable fixed points (or, more generally, multiple
attractors). In such situations, fluctuations may eventually push the
system from the vicinity of one such attractor to another, and the way
this occurs is typically not captured by the CLT, but rather by the
LDP. The aim of the present paper is to analyze in detail the
structure of the LDP for a specific class of fast-slow systems and
thereby provide concrete tools to characterize rare events as well as
noise induced transitions in such systems. We note that this question
has been investigated by many authors (see
e.g.~\cite{freidlin1978averaging,kifer1992averaging,liptser1996large,%
  veretennikov2000large,kifer2004averaging}), but few concrete results
exist that permit to actually compute the Hamiltonian associated with
the LDP: our goal here is to perform such calculations explicitly in
simple examples or indicate how they could be performed numerically in
more complicated ones. The examples we deal with have the peculiarity
that, on the one hand the fast variables evolve linearly once the slow
variable is held fixed, and on the other hand the fast variables act
on the slow variable through a quadratic nonlinearity. Such a
structure is very common in many examples related to the kinetic
theories of both turbulent flows and plasma physics. It appears
naturally for instance for the kinetic theory of the 2D Navier-Stokes
or the family of quasigeostrophic models (see
e.g.~\cite{bouchet2013kinetic} and references therein) or the kinetic
theory of plasma physics leading to either the Lenard-Balescu or the
Vlasov equation (see e.g.~\cite{nicholson1983introduction}), or more
generally the kinetic theory of systems with long range interactions
(see e.g.~\cite{campa2014physics}).

\subsection{Set-up}
\label{sec:setup}
We will consider fast-slow systems of the type
\begin{equation}
  \label{eq:19}
  \begin{cases}
     \dot X = f(X,Y)\\[8pt]
     \displaystyle d Y = \frac1{\alpha} b(X,Y) dt 
     + \frac1{\sqrt{\alpha}}\sigma(X,Y) dW(t)
  \end{cases}
\end{equation}
Here $f:\RR^{m\times n} \to \RR^m$ and $b:\RR^{m\times n} \to \RR^n$
are vector fields, $W(t)\in \RR^p$ is a standard $p$-dimensional
Wiener process, $\sigma: \RR^{m\times n} \to \RR^{n\times p}$, and
$\alpha>0$ is a parameter whose smallness measures the separation of
time scale between the slow $X\in \RR^m$ and the fast $Y\in \RR^n$.
To analyze the behavior of the slow variables $X$ when $\alpha\ll1$,
let us introduce the virtual fast process
\begin{equation}
  \label{eq:22}
  d \tilde Y_x = b(x,\tilde Y_x) d\tau  
     + \sigma(x,\tilde Y_x) dW(\tau)
\end{equation}
where $x$ is fixed. This equation is obtained from the equation for
$Y$ in~\eqref{eq:19} by setting $X$ to the fixed value $x$ and
rescaling time to the natural time scale of the fast~$Y$,
$\tau = t/\alpha$. Assume that the virtual fast process is ergodic at
every $x$ with respect to the invariant measure $\mu_x (dy)$ (which
may depend parametrically on $x$) and that the following expectation
exists
\begin{equation}
  \label{eq:23}
  \begin{aligned}
    F(x) & = \int_{\RR^n} f(x,y) \mu_x (dy) \equiv (Pf)(x)\\
    & = \lim_{T\to\infty} \frac1T \int_0^T f(x,\tilde Y^x(\tau)) d\tau
  \end{aligned}
\end{equation}
where the second equality follows from ergodicity.  Then for any
$\eps>0$ and any fixed $T<\infty$ we have
\begin{equation}
  \label{eq:24}
  \lim_{\alpha\to 0} \PP^x \left(\sup_{0\le t \le T} |X(t) - \bar X(t)|
    < \eps\right) = 1
\end{equation}
where $\PP^x$ denotes the probability conditional on $X(0)= \bar X(0)
= x$ and $\bar X$ satisfies the limiting equation
\begin{equation}
  \label{eq:25}
  \Dot{\bar X} = F(\bar X) 
\end{equation}
Property~\eqref{eq:24} is an instance of the LLN in the present
context.  For the reader's convenience, and also because the
derivation of~\eqref{eq:25} involve formal asymptotic tools that we
will need below, we recall this derivation in
Appendix~\ref{sec:llnderiv}.

The limiting equation in~\eqref{eq:25} is deterministic because we
assumed that there is no explicit noise acting on the slow variables
$X$ in~\eqref{eq:19} and we let $\alpha\to0$. In this limit the effect
of the fast~$Y$ on the slow~$X$ completely averages out. For small but
finite $\alpha$, however, the slow variables are subject to
fluctuations above their mean. To leading order, these fluctuations
can be captured by the Central Limit Theorem (CLT). More precisely,
denote by $C_{\!\tilde f}(x,\tau)$ the time-correlation matrix of
$\tilde f(x,y) \equiv f(x,y) - F(x)$ along the virtual fast process
$\tilde Y_x$,
\begin{equation}
  \label{eq:78}
  C_{\!\tilde f} (x,\tau)= \int _{\RR^n} \EE^y \left(\tilde f(x,\tilde
    Y_x(\tau)) \tilde f^{\,T}(x,y) + \tilde f(x,y) 
    \tilde f^{\,T} (x,\tilde Y_x(\tau))\right) \mu_x (dy)
\end{equation}
where $\EE^y$ denotes expectation over the virtual fast process
conditional on $Y^x(0)=y$. Assume that the following integral
exists
\begin{equation}
  \label{eq:43}
  \begin{aligned}
    A(x) & = \int_0^\infty\!\! C_{\!\tilde f} (x,\tau) d\tau,\\
    & = \lim_{T\to\infty} \frac1T \int_0^T d\tau \int_0^T d\tau'
    \tilde f(x,\tilde Y_x(\tau)) \tilde f^{\,T} (x,\tilde Y_x(\tau'))
  \end{aligned}
\end{equation}
and the following expectation exists
\begin{equation}
  \label{eq:41}
  \begin{aligned}
    B(x) & = \int_{\RR^n} \partial_x f(x,y) \mu_x(dy)\\
    & \quad + \int_0^\infty d\tau \int_{\RR^n} \left(\partial_y
    \EE^y f(x,\tilde Y^x(\tau))\right) \partial_x b(x,y) \mu_x(dy) 
  \end{aligned}
\end{equation}
where $(\partial_y f)_{i,j} = \partial f_i/\partial y_j$ and
$(\partial_x b)_{i,j} = \partial b_i/\partial x_j$.  Then on any
interval $t\in[0,T]$ with $T<\infty$, the process
\begin{equation}
  \label{eq:27}
  \tilde \xi = \frac{X - \bar X}{\sqrt{\alpha}}
\end{equation}
converges in distribution towards the Gaussian process solution of
\begin{equation}
  \label{eq:33}
  d \xi = B(\bar X) \xi dt + \eta(\bar X) dW(t)
\end{equation}
where $\eta(x)$ is a $m\times m$ matrix such that
$(\eta \eta^T)(x)= A(x)$ and $W(t)$ is a $m$-dimensional Wiener
process -- the derivation of~\eqref{eq:33} via formal asymptotic
expansion techniques is recalled in Appendix~\ref{sec:cltderiv}.

While the CLT indicates that typical fluctuations of the slow
variables around their mean are of order $O(\sqrt{\alpha})$ on
time-scales that are $O(1)$ in $\alpha$, it does not permit to
estimate the probability of deviations of order one away from this
mean. On these time-scales, large deviations are expected to be rare,
and their probability can be estimated by a LDP which takes the
following form. Suppose that the following expectation over the
virtual fast process exists
\begin{equation}
  \label{eq:6}
  H(x,\theta) = \lim_{T\to\infty} \frac1T \log \EE^y
  \exp\left(\theta \cdot \textstyle \int_0^T f(x,\tilde Y_x(\tau)) d\tau\right)
\end{equation}
and define the Lagrangian associated with this Hamiltonian
\begin{equation}
  \label{eq:7}
  \mathcal{L}(x,y) = \sup_{\theta} \left(y\cdot x-H(x,\theta)\right)
\end{equation}
as well as the action
\begin{equation}
  \label{eq:8}
  I_T(x) = \int_0^T \mathcal{L}(x(t),\dot x(t))dt
\end{equation}
Then this action permits to estimate the probability that the slow
process wanders away from its mean behavior in the sense that for any
$\Gamma\subset \{\gamma \in C([0,T],\RR^m): \gamma(0) =x\}$ the
following LDP holds
\begin{equation}
  \label{eq:9}
  \begin{aligned}
    -\inf_{\gamma\in \Gamma^\circ} I_T(\gamma) & \le \liminf_{\alpha\to0} \alpha
    \log \PP( X\in \Gamma) \\
    & \le \limsup_{\alpha\to0} \alpha \log \PP(
    X\in \Gamma) \le - \inf_{\gamma\in \bar \Gamma} I_T(\gamma)
  \end{aligned}
\end{equation}
The action~\eqref{eq:8} is also useful if one is interested in the
slow motions on longer time-scales that can be $O(\alpha^{-1})$ or
even $O(\exp(C/\alpha))$ for $C>0$.  On these time-scales, large
deviations stop being rare and may, for example, lead to random
transitions between the different attractors of the limiting equation
in~\eqref{eq:25} if there are more than one. For example suppose that
$D\subset \RR^m$ is an open set that contains a single stable
attracting point $x_D$ of the limiting equation~\eqref{eq:25} and let
us consider the first exit time from $D$
\begin{equation}
  \label{eq:64}
  T_D = \inf\{t>0: X(t) \not \in D\}
\end{equation}
Then for any $x\in D$ we have
\begin{equation}
  \label{eq:65}
  \lim_{\alpha\to0} \alpha \log \EE^x T_D = \inf_{y\in \partial D} V(x_D,y)
\end{equation}
where $V(x,y)$ is the quasipotential
\begin{equation}
  \label{eq:66}
  V(x,y) = \inf_{T>0} \inf_{\substack{\gamma(0)=x\\ \gamma(T) = y}} I_T(\gamma)
\end{equation}

While the results above are well-known, in particular
expression~\eqref{eq:6} for the Hamiltonian in the LDP
(see~\cite{freidlin1978averaging,kifer1992averaging,%
  veretennikov2000large,kifer2004averaging}), little attention has
been given to the explicit form this LDP takes via calculation of the
Hamiltonian. As stated above, our main aim here is to provide tools to
perform such calculations, either analytically or numerically. Even if it
goes without saying, we should stress that the
Hamiltonian~\eqref{eq:6} cannot be deduced simply from the limiting
equations~\eqref{eq:25} from the LLN and~\eqref{eq:33} from the CLT.
Indeed, we may naively try to somehow recombine these two limiting
equation and write down a nonlinear SDE whose LLN and CLT are
precisely~\eqref{eq:25} and~\eqref{eq:33}: since the noise in this SDE
would be scaled by a factor $\sqrt{\alpha}$ one may be tempted to
think that its associated LDP would also yield the LDP for the slow
variables solution of~\eqref{eq:19}. This program, however, is not
achievable in general because there is no SDE for the slow variable
alone that admits~\eqref{eq:25} as LLN and \eqref{eq:27} as CLT,
except in special cases. Indeed the drift term in this equation would
have to be $F(X)$ to leading order in $\alpha$, and setting $X = \bar
X + \sqrt{\alpha}\, \xi$ and expanding in $\alpha$ would give a linear
drift in the CLT equation equivalent to~\eqref{eq:33} involving
$\partial_x F(\bar X)$. And there lies the problem: $B(x) \not=
\partial_x F(x)$ in general.

As we will see below, this problem is consistent with the fact that
the LDP for the slow variables~$X$ in~\eqref{eq:19} is different in
form to that of an SDE for $X$ alone -- in particular, the
Hamiltonian~\eqref{eq:6} involved in this LDP is typically
non-quadratic in the momenta conjugate to~$X$, unlike that of an SDE
with small noise. This essentially means that the effect of the noise
induced by the fast variables on the slow ones cannot be modeled as a
Gaussian white-noise in general. Below, we will re-derive the LDP
stated above using formal asymptotic expansions tools. These results
are complementary to the rigorous results proven
e.g.
in~\cite{freidlin1978averaging,kifer1992averaging,liptser1996large,%
  veretennikov2000large,kifer2004averaging}. In particular we obtain
an equation for the Hamiltonian which, to the best of our knowledge,
is new and can be solved explicitly in some nontrivial examples.

\subsection{Organization}
\label{sec':organization}
The remainder of this paper is organized as follows. In
Sec.~\ref{sec:LDP} we derive the LDP for the fast-slow
system~\eqref{eq:19}. This derivation is formal, and to connect it
with the results stated above we put it within the context of
Donsker-Varadhan theory (Sec.~\ref{sec:DVLDT}) of large deviation and
Gartner-Ellis theorem (Sec.~\ref{sec:gartnerellis}). We also discuss
in Sec.~\ref{sec:link} the link between the LLN and the LDP on the one
hand, and the CLT and the LDP on the other. In
Sec.~\ref{sec:specific}, we specialize the LPD to a class of fast-slow
system for the which the equation for the Hamiltonian can be
simplified. In Sec.~\ref{sec:study} we use these results to study a
test-case example. Finally, some concluding remarks are given in
Sec.~\ref{sec:homog}.

\section{Derivation of the Large Deviation Principle (LDP)}
\label{sec:LDP}

The slow-fast system in~\eqref{eq:19} defines a Markov process with
generator $L = L_0 + \alpha^{-1} L_1$ where
\begin{equation}
  \label{eq:26}
  \begin{aligned}
    L_0 &= f(x,y) \cdot \partial_x,\\
    L_1 &= b(x,y) \cdot \partial_y + \tfrac12 a(x,y) : \partial_y \partial_y 
  \end{aligned}
\end{equation}
with $a(x,y) = (\sigma\sigma^T)(x,y)$, and
$a(x,y) : \partial_y \partial_y$ is the contraction of the two tensors
$a$ and $\partial_y \partial_y$ (i.e. the trace of the product
$a \partial_y \partial_y$). Given any suitable test
function~$\phi: \RR^m \to \RR$ the expectation
\begin{equation}
  \label{eq:29}
  u(t,x,y) = \EE^{x,y} \phi(X(t))
\end{equation}
therefore satisfies the backward Kolmogorov equation
\begin{equation}
  \label{eq:28}
  \partial_t u = L_0 u + \frac1{\alpha} L_1 u, \qquad u(0) = \phi
\end{equation}
Our derivation of the LDP for the slow variables $X$ in~\eqref{eq:19}
is based on formal asymptotic analysis of this equation.  This
approach is also at the core of the formal derivations of the LLN
equation in~\eqref{eq:25} presented in Appendix~\ref{sec:llnderiv} and
the CLT equation in~\eqref{eq:33} presented in
Appendix~\ref{sec:cltderiv}.

Since the LDP for the slow variables $X$ in~\eqref{eq:19} is concerned
with estimating the probability of having a fluctuation of $X$ away
from $\bar X$ that is $O(1)$ in $\alpha$, and since such probability
is expected to be exponentially small in $\alpha^{-1}$ on time-scales
that are $O(1)$ in $\alpha$, let us consider the expectation
\begin{equation}
  \label{eq:11}
  u(t,x,y) = \EE^{x,y} \exp\left(\frac1\alpha h(X(t)))\right),
\end{equation}
so that $u$ satisfies~\eqref{eq:28} for the initial condition 
\begin{equation}
  \label{eq:10}
  u(0,x,y) = \exp\left(\frac1\alpha h(x)\right),
\end{equation}
and look for a solution of the type
\begin{equation}
  \label{eq:36}
  u(t,x,y) = \Big(w(t,x,y) + O(\alpha)\Big) 
  \exp\left(\frac1\alpha S(t,x,y)\right)
\end{equation}
where both $S$ and $w$ are assumed to be independent of
$\alpha$. Inserting this ansatz in~\eqref{eq:28} and collecting terms
of increasing power in $\alpha$, we obtain at leading order,
$O(\alpha^{-2})$:
\begin{equation}
  \label{eq:62}
  \tfrac12 a : \partial_y S\partial_y S =0 
\end{equation}
This equation indicates that $S$ depends on $x$ but not $y$,
$S(t,x,y) \equiv S(t,x)$, or, equivalently,
\begin{equation}
  \label{eq:63}
  PS = S
\end{equation}
where $P$ is the expectation operator with respect to the invariant
measure of the fast virtual process~\eqref{eq:22}, see~\eqref{eq:23}.
The function $S$ plays the role of the action of the LDP, since
combining~\eqref{eq:11} and~\eqref{eq:36} with
$S(t,x,y) \equiv S(t,x)$ implies that
\begin{equation}
  \label{eq:80}
  S(x,t) = 
  \lim_{\alpha\to0} \alpha \log \EE^{x,y} \exp\left(\frac1\alpha h(X(t)))\right)
\end{equation}
which is a variant of Varadhan's Lemma~\cite{dembo2009large}. The
question which remains to be addressed is how to estimate $S(x,t)$?

To this end, go to next order, $O(\alpha^{-1})$, where the
ansatz~\eqref{eq:36} used in~\eqref{eq:28} gives
\begin{equation}
  \label{eq:28ccc}
    w \partial_t S  = w L_0 S +  L_1 w.
\end{equation}
where we used the properties that $L_0$ is a first order operator,
i.e. $L_0(fg) = f L_0 g + g L_0 f$, and $\partial_y S = L_1 S = 0$.
Since $S = P S$ from~\eqref{eq:63}, we can get an equation for $S$
from~\eqref{eq:28ccc} by dividing this equation by $w$ (note that
$w>0$ since $u>0$ by definition, see~\eqref{eq:11}) and applying
$P$. This gives
\begin{equation}
  \label{eq:44}
    \begin{aligned}
      \partial_t S  &= L_0 S + w^{-1} L_1 w\\
      & = P L_0 P S + P(w^{-1} L_1 w)
  \end{aligned}
\end{equation}
Inserting~\eqref{eq:44} back in~\eqref{eq:28ccc} gives
\begin{equation}
  \label{eq:46}
  L_1 w + L_0 S w =  \left(PL_0 P S + P (w^{-1} L_1 w)
  \right) w
\end{equation}
This equation can be written explicitly as
\begin{equation}
  \label{eq:47}
  b(x,y) \cdot \partial_y w + \tfrac12 a(x,y) : \partial_y \partial_y
  w  + f(x,y) \cdot \partial_x S\ w
  = H(x,\partial_xS) w
\end{equation}
where we defined 
\begin{equation}
  \label{eq:48}
  H(x,\partial_x S) = F(x) \cdot \partial_x S + \tfrac12\int_{\RR^n} \mu_x(dy) \, 
  a(x,y) : \partial_y \log w\,  \partial_y \log w 
\end{equation}
Here we used $w^{-1} L_1 w= L_1 \log w + \tfrac12 a(x,y) : \partial_y
\log w\, \partial_y \log w $ and $PL_1 = 0$, to express the last term
at the right hand side of~\eqref{eq:44} as
\begin{equation}
  \label{eq:44b}
  P (w^{-1} L_1 w)  = \tfrac12
  P (a(x,y) : \partial_y \log w\,  \partial_y \log w )
\end{equation}
Since $x$ and $\partial_x S$ only enter as parameters
in~\eqref{eq:47}, this equation can be viewed an eigenvalue problem
for the operator~$L_1$, where
$H(x,\partial_x S) - f(x,y)\cdot \partial_x S $ plays the role of
(leading) eigenvalue. The function $H(x,\partial_x S)$ also is the
Hamiltonian of the LDP we are trying to derive. Indeed, if we go back
to equation~\eqref{eq:44} for $S$, we see that it can be expressed in
terms of $H(x,\partial_xS)$ as the following Hamilton-Jacobi equation:
\begin{equation}
  \label{eq:49}
  \partial_t S = H(x,\partial_x S), \qquad S(x,0) = h(x)
\end{equation}
This also means that if we introduce the Lagrangian associated with
$H$ via
\begin{equation}
  \label{eq:50}
  \mathcal{L}(x,y) = \sup_{\theta}\left( y \cdot \theta - H(x,\theta) \right)
\end{equation}
then the process $X$ will satisfy a large deviation principle with
respect to the action
\begin{equation}
  \label{eq:51}
  I_T(x) = \int_0^T \mathcal{L}(x(t),\dot x(t)) dt
\end{equation}
and the solution to~\eqref{eq:49} can be expressed as
\begin{equation}
  \label{eq:81}
  S(t,x) = \inf\{ h + I_t(\tilde x)\} 
\end{equation}
where the infimum is taken over all paths~$\tilde x: [0,t) \to \RR^m$
such that $\tilde x(t) = x$. Since $S(t,x)$ permits via~\eqref{eq:80}
to evaluate the limit of $\alpha$ time the logarithm of the expectation
in~\eqref{eq:80}, this is again Varadhan's Lemma. 

To make these results concrete, it remains to see whether we can
analyze~\eqref{eq:47} and get a more convenient equation for
$H(x,\theta)$. This will be done in Sec.~\ref{sec:specific} for a
specific class of systems. Before going there, however, we show that
our results are consistent with those of Donsker-Varadhan large
deviations theory (Sec.~\ref{sec:DVLDT}) as well as with Ellis-Gartner
theorem (Sec.~\ref{sec:gartnerellis}). We also discuss the link
between the LDP we just derived and the LLN on the one hand and the CLT
on the other (Sec.~\ref{sec:link}).

\subsection{Connection with Donsker-Varadhan Theory of Large
  Deviations}
\label{sec:DVLDT}

Denoting the solution of~\eqref{eq:47} for~$x$ and
$\partial_x S= \theta$ fixed by $w(y,x,\theta)$, this solution can be
expressed as the expectation
\begin{equation}
  \label{eq:45}
  \begin{aligned}
    w(y,x,\theta) & = \lim_{T\to\infty} \EE^y \exp\Bigl(-T H(x,\theta) +
    \theta\cdot\int_0^T f(x,\tilde Y_x (\tau))d\tau \Bigr)\\
    & = \lim_{T\to\infty} e^{-T H(x,\theta)} \EE^y \exp\Bigl(
    \theta\cdot\int_0^T  f(x,\tilde Y_x (\tau))d\tau \Bigr)
  \end{aligned}
\end{equation}
where $\tilde Y_x (t)$ is the virtual fast process, solution
to~\eqref{eq:22}, and $H(x,\theta)$ needs to be adjusted to make the
limit converge to a finite, nonzero value. The requirement that such
an adjustment be possible also gives the condition for the solution
to~\eqref{eq:47} to exist and be uniquely given by~\eqref{eq:45}. By
taking the logarithm of the factor under the limit in~\eqref{eq:45},
it is easy to see that this requirement imposes as a necessary
condition that the following limit exists
\begin{equation}
  \label{eq:76}
  H(x,\theta) = \lim_{T\to\infty} \frac1T \log \EE^y \exp\Bigl(
  \theta\cdot\int_0^T  f(x,\tilde Y_x (\tau))d\tau \Bigr)
\end{equation}
This is the alternative expression for $H$ already given
in~\eqref{eq:6} that is consistent with
the one in Donsker-Varadhan large deviations theory.

\subsection{Connection with Gartner-Ellis Theorem} 
\label{sec:gartnerellis}

Gartner-Ellis Theorem tells us that the Hamiltonian in~\eqref{eq:48}
can also be defined by taking the limit as $\alpha\to0$ of
\begin{equation}
  \label{eq:52}
  \alpha \log \EE \exp\left( \frac1\alpha \int_0^T \theta(t) \cdot f(x(t),
     Y_x (t))dt\right) 
\end{equation}
where $ Y_x(t)$ is the solution of the second equation in~\eqref{eq:19}
for a given $x(\cdot)$, i.e.
\begin{equation}
  \label{eq:53}
  d Y_x = \frac1\alpha b(x(t),Y_x) dt 
     + \frac1{\sqrt{\alpha}}\sigma(x(t),Y_x) dW(t)
\end{equation}
Letting
\begin{equation}
  \label{eq:54}
  \hat u(t,y) = \EE^y \exp\left( \frac1\alpha \int_0^t \theta(t') \cdot F(x(t'),
    Y_x (t'))dt'\right)
\end{equation}
this function satisfies
\begin{equation}
  \label{eq:55}
  \partial_t \hat u = \frac1\alpha L_1 \hat u 
  + \frac1\alpha \theta(t) \cdot f(x(t),
     y)\, \hat u
\end{equation}
Look for a solution of the type
\begin{equation}
  \label{eq:56}
  \hat u(t,y) = \hat w(t,y) \exp\left(\frac1\alpha \hat \phi(t)\right)
\end{equation}
Proceeding similarly as above then lead to the following system of
equations for $\hat \phi$ and $\hat w$ (compare~\eqref{eq:47}
and~\eqref{eq:49}):
\begin{equation}
  \label{eq:47b}
  b(x,y) \cdot \partial_y \hat w + \tfrac12 a(x,y) : \partial_y \partial_y
  \hat w  + f(x,y) \cdot \theta \ \hat w
  = H(x,\theta) \hat w
\end{equation}
and
\begin{equation}
  \label{eq:49b}
  \partial_t \hat \phi = H(x,\theta)
\end{equation}
for the Hamiltonian $H$ defined in~\eqref{eq:48}. This also means that
\begin{equation}
  \label{eq:57}
  \lim_{\alpha\to0}
  \alpha \log \EE \exp\left( \frac1\alpha \int_0^T \theta(t) \cdot f(x(t),
     Y_x (t))dt\right) = \int_0^T H(x(t),\theta(t))dt
\end{equation}
which leads again to the action in~\eqref{eq:51}.

\subsection{Link between the LDP, the LLN, and the CLT}
\label{sec:link}

Suppose that we expand the exponential in \eqref{eq:76} to second order
in $\theta$, take the expectation, then expand the logarithm to second
order in $\theta$ as well, and finally take the limit as
$T\to\infty$. This sequence of operations corresponds to making a
cumulant expansion of the variable
$\theta\cdot\int_0^T f(x,\tilde Y_x (\tau))d\tau$ truncated to second
order, and it gives the following quadratic approximation for~$H$:
\begin{equation}
  \label{eq:77}
  H_\text{quad}(x,\theta) = \theta\cdot F(x) + \tfrac 12 \theta^T A(x) \theta
\end{equation}
where $F(x)$ is defined in~\eqref{eq:23} and $A(x)$
in~\eqref{eq:43}. This is the Hamiltonian for the LDP associated with the SDE
\begin{equation}
  \label{eq:79}
   dX = F(X) dt + \sqrt{\alpha}\, \eta(X) dW(t)
\end{equation}
in the limit as $\alpha\to0$. The process defined by~\eqref{eq:79}
satisfies a LLN principle with limiting equation~\eqref{eq:25},
meaning that the LDP contains the information about the LLN. Yet, it
also highlights the subtle (and well-know) differences between the CLT
and the LDP. Indeed, the process defined by~\eqref{eq:79} satisfies a
CLT with respect to
\begin{equation}
  \label{eq:82}
  d\xi = \partial_x  F(\bar X ) \xi dt + \eta(\bar X) dW(t) 
\end{equation}
This equation is not identical with~\eqref{eq:33} -- their drift terms
are different. In fact, it is easy to see that we would have to add
deterministic terms of order $O(\sqrt{\alpha})$ in~\eqref{eq:79} in
order that the CLT associated with this modified equation coincide
with~\eqref{eq:33} -- as already mentioned in the introduction, the
addition of these terms would render the resulting equation unclosed
since they depend on $\xi$ rather than $X$.  These additional terms do
not affect the Hamiltonian of the LDP which would still
be~\eqref{eq:77}. This goes to show that the CLT cannot be recovered
from the LDP: in the range of values for $X$ where it applies, it
contains finer information than that in the LDP.

Conversely, the LDP cannot be deduced from the CLT, as the actual $H$
in~\eqref{eq:76} is different in general from its quadratic
approximation $H_\text{quad}$. More precisely, $H=H_\text{quad}$
\textit{iff} $\theta\cdot\int_0^T f(x,\tilde Y_x (\tau))d\tau$ is
Gaussian, and Marcinkiewicz's theorem states that in all other cases
$H$ is not a polynomial of $\theta$ of any order (i.e. its expansion
in $\theta$ involves infinitely many terms).  As we will see below in
Sec.~\ref{sec:specific}, $H_\text{quad}=H$ if the fast~$Y$ enter
linearly the equation for the slow~$X$.  When this is not the case and
$H_\text{quad}\not=H$, $H_\text{quad}$ can be used to describe
moderate fluctuations of order $O(\alpha^\nu)$ with
$\frac12\le \nu < 1$, that is, outside the range of validity of the
CLT but only moderately so. To describe large fluctuations of order
$O(1)$, however, we need to use the actual $H$ in~\eqref{eq:76}.

\section{A Specific Class of Systems}
\label{sec:specific}

Next let us specialize~\eqref{eq:19} to systems in which the dynamics
of the fast~$Y$ is linear and they enter quadratically the equation
for the slow~$X$
\begin{equation}
  \label{eq:10bbb}
  \begin{cases}
    \begin{aligned}
      \dot X &= r(X) + Y^T s(X) + Y^T M(X) Y \\
      dY &= -\frac1\alpha L(X) Y + \frac1{\sqrt{\alpha}}\sigma(X) dW
    \end{aligned}
  \end{cases}
\end{equation}
where $X\in \RR$ (the generalization to the vectorial case is
straightforward but it makes notations more cumbersome, so we will
stick to the scalar case here -- see however Sec.~\ref{sec:2D} for an
illustration with $X\in \RR^2$), $Y\in \RR^n$, $r:\RR\to \RR$ (e.g.
$r(x) = - \nu x$ for some $\nu>0$), $s: \RR \to \RR^n$,
$L: \RR\to \RR^{n\times n}$ is a positive-definite matrix,
$M : \RR \to \RR^{n\times n}$ is a symmetric matrix, and
$\sigma: \RR \to\RR^{n\times n}$.  For such systems, the
equation~\eqref{eq:47} leading to the Hamiltonian of the LDP can be
written down more explicitly.

To see how, notice first that the virtual fast process defined
in~\eqref{eq:22} is given explicitly by
\begin{equation}
  \label{eq:83}
  \tilde Y^x(\tau) = e^{-L(x) \tau} y + \int_0^\tau  e^{-L(x)
    (\tau-\tau')} \sigma(x) dW(\tau')
\end{equation}
As a result
\begin{equation}
  \label{eq:11bb}
  \mu_x (dy) = (2\pi)^{-n/2} (\det C(x))^{-1/2} \exp\left(-\frac12 y^T C^{-1}(x) y\right) dy
\end{equation}
where $C(x)$ is the equilibrium covariance matrix of $\tilde
Y^x(\tau)$ satisfying the Lyapunov equation
\begin{equation}
  \label{eq:12}
  L(x) C(x) + C(x) L^T(x) =  a(x)
\end{equation}
where $a(x) = (\sigma\sigma^T)(x)$. This means that the limiting
equation~\eqref{eq:25} from the LLN reads
\begin{equation}
  \label{eq:13}
  \Dot{\bar X} = r(\bar X) + \text{tr} \left(C(\Bar X) M(\bar X)\right)
\end{equation}
and the equation \eqref{eq:33}  from the CLT
reads
\begin{equation}
  \label{eq:14}
  d\xi = r'(\bar
  X)\xi  dt + \text{tr} \left(C(\Bar X) M'(\bar X)\right) \xi dt +
  g(\bar X) \xi dt+\eta(\bar X) dW(t)
\end{equation}
Here
\begin{equation}
  \label{eq:84}
  g(x) = -\int_0^\infty \text{tr} \left(
    C(x) (L'(x) +  [L'(x)]^T ) e^{-L^T(x)\tau} M(x) e^{-L(x)\tau} \right) d\tau
\end{equation}
and
\begin{equation}
  \label{eq:15}
  \begin{aligned}
    \eta^2(x) & = \int_0^\infty s^T(x) \left(C(x) e^{-L^T(x)\tau}+
      e^{-L(x)\tau} C(x) \right)
    s(x) d\tau \\
    & \quad + 4\int_0^\infty \text{tr}\left(C(x) e^{-L(x)\tau} M(x)
      e^{-L^T(x)\tau} C(x)M(x)\right)
    d\tau\\
    & = s^T(x) L^{-1}(x) a(x) L^{-T}(x)  s(x)  \\
    & \quad + 4\int_0^\infty \text{tr}\left(C(x) e^{-L(x)\tau} M(x)
      e^{-L^T(x)\tau} C(x)M(x)\right)
    d\tau\\
  \end{aligned}
\end{equation}
where we used the Lyapunov equation~\eqref{eq:12}.

Turning ourselves to the LDP next,~\eqref{eq:47} is explicitly
\begin{equation}
  \label{eq:47s}
  \begin{aligned}
    & -L(x)y \cdot \partial_y w + \tfrac12 a(x) : \partial_y \partial_y
    w\\
    & \quad + \left(r(x) + y ^T\cdot s(x) + y^T M (x) y \right) \theta\
    w 
    = H(x,\theta) w
  \end{aligned}
\end{equation}
where $\theta=\partial_x S$ and
\begin{equation}
  \label{eq:48s}
  \begin{aligned}
    H(x,\theta) & = \left(r(x)+ \text{tr} (C(x) M(x)) 
      \right)\theta
      \\
      & \quad + \tfrac12\int \mu_x(dy) \, a(x) : \partial_y \log
      w\, \partial_y \log w
    \end{aligned}
\end{equation}
Look for a solution of the form
\begin{equation}
  \label{eq:17}
  w = \exp(y^T m(x,\theta) +  y^T N(x,\theta) y)
\end{equation}
for some unknown $m(x,\theta)$ and $N(x,\theta)$. Then (dropping the
dependencies in $x$ for simplicity of notation)
\begin{equation}
  \label{eq:20}
  H(\theta) = \left(r+\text{tr}(C M) 
      \right)\theta  + \tfrac12 m^T(\theta) a m(\theta) 
      +  2\,\text{tr} \left (C N(\theta) a N(\theta)\right)
\end{equation}
and~\eqref{eq:47s} becomes
\begin{equation}
  \label{eq:21}
  \begin{aligned}
    &-y^T N(\theta) L y - y^T L^T N(\theta) y - y^T L^T m(\theta)+
    \text{tr}(a N(\theta))  \\
    &+ 2 y^T N(\theta) a
    m(\theta) + 2 y^T N(\theta) a N(\theta)y\\
    &+ (y^T s + y^T M y) \theta = \text{tr}(C M) \theta  
      +  2\,\text{tr} \left (a N(\theta)C N(\theta)\right)
  \end{aligned}
\end{equation}
Collecting the terms that are of order 0, 1, and 2 in $y$,
respectively, gives the equations
\begin{equation}
  \label{eq:74}
  \text{tr}(a N(\theta)) =
  \text{tr}(C M) \theta + 2 \, \text{tr} \left (C
    N(\theta) a N(\theta)\right)
\end{equation}
\begin{equation}
  \label{eq:75}
  (L^T-2N(\theta) a)  m(\theta) = s\, \theta
\end{equation}
\begin{equation}
  \label{eq:73}
   N(\theta) L +   L^T N(\theta) 
  = 2 N(\theta) a N(\theta) + M \theta 
\end{equation}
By right multiplying~\eqref{eq:73} by $C$, taking the trace, and using
the Lyapunov equation~\eqref{eq:12}, it is easy to see that the result
is~\eqref{eq:74}, i.e. if \eqref{eq:73} is satisfied
then~\eqref{eq:74} automatically holds. We can also
solve~\eqref{eq:75} in $m(\theta)$ to get
\begin{equation}
  \label{eq:16}
  m(\theta) = (L^T-2N(\theta) a)^{-1} s \, \theta
\end{equation}
and be left with solving \eqref{eq:73} in $N(\theta)$ -- note that
this equation may have more than one solution, and we should take the
one such that $N(0) = 0$, so that $H(0)=0$.  Inserting~\eqref{eq:74}
and \eqref{eq:16} in~\eqref{eq:20} we then obtain the Hamiltonian of
the LDP in terms of $N(\theta)$ alone
\begin{equation}
  \label{eq:20bb}
    H(\theta)  = r\theta + \text{tr}
    \left( a N(\theta)\right) 
     + \tfrac12 s^T (L^T-2N(\theta) a)^{-T} a 
    (L^T-2N(\theta) a)^{-1} s \, \theta^2
\end{equation}

The solution to~\eqref{eq:73} is not available explicitly
in general. There is one trivial case, however, namely when $M=0$. In
this case it is easy to see that $N(\theta) =0$, $m(\theta)= L^{-T} s
\, \theta$ and the Hamiltonian is quadratic
\begin{equation}
  \label{eq:85}
  H(\theta) = r \theta 
     + \tfrac12 s^T L^{-1} a L^{-T} s \, \theta^2
\end{equation}
This Hamiltonian is that of the LDP  associated with SDE
\begin{equation}
  \label{eq:86}
  d X = r(X)  dt +\sqrt{\alpha} \eta_0(X) dW(t)
\end{equation}
where $\eta_0(x)$ is the factor defined in~\eqref{eq:15} evaluated at
$M=0$. It is easy to see that this the limiting equation from the LLN
for this equation is~\eqref{eq:13} (with $M=0$) and the equation from
the CLT is~\eqref{eq:14} (again with $M=0$).  Thus, if the fast
variables are Gaussian, and their action on the slow one is linear,
the LDP contains all the information about the CLT, and the
SDE~\eqref{eq:86} can be used to investigate large deviations. Notice
that this includes nontrivial situations with metastability, when
$\dot x = r(x)$ has more than one stable fixed point and one is
interested in the rate and mechanism of
transitions between these points. 

Another case where the Hamiltonian can be computed explicitly is the
following one: assume that $a$ is invertible, and that the following
conditions hold:
\begin{equation}
La = aL^T \quad \text{and} \quad L^TMa = MaL^T.
\label{eq:conditions-solution-N}
\end{equation}
Then, it is straightforward to check that whenever
$B(\theta) = \left(L^T\right)^2 - 2\theta Ma$ admits a square root,
the matrix
\begin{equation}
N(\theta) = \frac{1}{2} \left[ L^T - \sqrt{B(\theta)} \right] a^{-1}
\label{eq:solution-N}
\end{equation}
satisfies $N(\theta)L = L^T N(\theta)$, and is a solution of
\eqref{eq:73}. Using \eqref{eq:conditions-solution-N}, we also have
that $aB(\theta)$ is symmetric, so inverting \eqref{eq:solution-N} and
using \eqref{eq:73}, we prove that $N(\theta)$ is symmetric, which is consistent with its definition \eqref{eq:17}. It is then also easy to
prove that $a\sqrt{B(\theta)}^{-1}=\sqrt{B^T(\theta)}^{-1}a$, so the
Hamiltonian \eqref{eq:20bb} reads
\begin{equation}
  \label{eq:solution-H}
    H(\theta)  = r\theta + \tfrac 12 \text{tr}
    \left(L - \sqrt{L^2 - 2\theta aM} \right) 
     + \tfrac12 s^T \left[L^2 - 2\theta aM\right]^{-1} a 
     s\,  \theta^2\,,
\end{equation}
whenever the square roots and inverses appearing in this equation
exist (which is the case for $\theta=0$). Note that the square root in
the trace should be chosen such that $H(\theta=0)=0$.

\section{A Case Study}  
\label{sec:study}

In this section, we illustrate our results on the following test case
example:
\begin{equation}
  \label{eq:1}
  \begin{cases}
    \displaystyle\dot X = \frac1K \sum_{k=1}^K Y_k^2 - \nu X&\\[8pt]
    \displaystyle dY_k = - \frac1{\alpha} \gamma(X) Y_k dt +
    \frac{\sigma}{\sqrt{\alpha}}
      \, dW_k, \qquad & k=1,\ldots, K
  \end{cases}
\end{equation}
where $W_k$ are independent Wiener processes, $\alpha>0$, $\nu>0$ and
$\sigma$ are parameters and $\gamma(X)>0$ is a function to be
specified later. \eqref{eq:1} consists of a scalar~$X$ (the vectorial
case is discussed below in Sec.~\ref{sec:2D}) coupled to $K$
Ornstein-Uhlenbeck processes $Y_k$ which each feel an independent
noise and whose common decay rate depends on $X$. \eqref{eq:1} is in
the class of~\eqref{eq:10bbb} with
\begin{equation}
  \label{eq:87}
  \begin{aligned}
    r(x) &= -\nu x, \quad s(x) = 0, \quad M(x) = K^{-1}\, \text{\it
      Id},\\
    L(x) &= \gamma(x)\,\text{\it Id}, \quad \sigma(x) = \sigma\,
    \text{\it Id}
  \end{aligned}
\end{equation}
We will be interested in studying~\eqref{eq:1} in the limits
$\alpha\to0$ and $K\to\infty$ -- the former limit is in the realm of
the formalism developed here, whereas the latter can be estimated by
direct calculation. As we will below these two limits commute.

\subsection{The limit as $\alpha\to0$}
\label{sec:alp0}

Using the formulas given in section~\ref{sec:specific}, it is easy to
see that the LLN equation~\eqref{eq:13} becomes
\begin{equation}
  \label{eq:58}
  \Dot{\bar X}= \frac{\sigma^2}{2\gamma(\bar X)} - \nu \bar X
\end{equation}
and the CLT equation~\eqref{eq:14} is
\begin{equation}
  \label{eq:59}
  d \xi = -\left(\nu +\frac{\sigma^2 
      \gamma'(\bar{X})}{2\gamma^2(\bar{X})} \right)\xi dt + 
  \frac{\sigma^2}{\sqrt{2K\gamma^{3}(\bar X)}} dW(t)
\end{equation}
As far as the LDP is concerned, note that \eqref{eq:1} is of the form
\begin{equation}
  \label{eq:Yk-independent}
  \dot{X} = \frac{1}{K}\sum_{k=1}^K f_0\left(X,Y_k\right) + f_1(X)
\end{equation}
where $Y_k$ are {\em i.i.d.} random processes. From \eqref{eq:76}, we
see that the Hamiltonian reads
\begin{equation}
  \label{eq:H-Yk-independent}
  H(x,\theta) =  \theta f_1(x) + K H_0\left(x,\frac{\theta}{K}\right)
\end{equation}
with $H_0$ the one-particle Hamiltonian
\begin{equation}
  \label{eq:H_0-Yk-independent}
  H_0(x,\theta) = \lim_{T\to\infty} \frac1T \log \EE^{y_1} \exp\Bigl(
  \theta\cdot\int_0^T  f_0(x,\tilde Y_{1,x} (\tau))d\tau \Bigr).
\end{equation}
To get the Lagrangian associated with this Hamiltonian, we solve
\begin{equation}
  \label{eq:69}
  \dot x = \frac{\partial H}{\partial \theta} = f_1(x)+
  \frac{\partial H_0}{\partial \theta} \left(x,\frac{\theta}{K}\right)
\end{equation}
and get $\theta/K$ as a function of $(x,\dot{x})$. Denoting this
solution as $\theta = K \vartheta(x,\dot x)$ and using
\eqref{eq:H-Yk-independent} and \eqref{eq:69}, we then deduce that the
Lagrangian is proportional to $K$ and reads
\begin{equation}
  \label{eq:L-Yk-independent}
  \begin{aligned}
    \mathcal{L}(x,\dot x) & = \dot x\theta - H(x,\theta)\\
    & = K\left( \vartheta(x,\dot x)\frac{\partial H_0}{\partial
        \theta} \left(x,\vartheta(x,\dot x)\right) -
      H_0\left(x,\vartheta(x,\dot x)\right) \right).
  \end{aligned}
\end{equation}
As a result, the path that minimizes the action \eqref{eq:51}
(instanton) does not depend on $K$, but the probabilistic weight of
this path decreases exponentially with $K$. The quasi-potential
$V(x)$, which by definition is the solution of
\begin{equation}
  \label{eq:91}
  H(x,\partial_x V) = 0
\end{equation}
with $H$ given by \eqref{eq:H-Yk-independent}, is then also
proportional to $K$.

Going back to the particular case \eqref{eq:1}, the one-particle Hamiltonian \eqref{eq:H_0-Yk-independent} can be computed using \eqref{eq:solution-H}. Indeed, we are now solving a one-dimensional problem, so the conditions \eqref{eq:conditions-solution-N} are fulfilled and \eqref{eq:solution-H} is simply
\begin{equation}
  \label{eq:H_0-case}
  H_0(x,\theta) = \frac12 \left[ \gamma(x) 
  -   \sqrt{\gamma^2(x) - 2
      \sigma^2\theta}\right],
\end{equation}
defined whenever $\theta\leqslant\gamma^2(x)/2\sigma^2$.
To get the Lagrangian associated with this Hamiltonian, we solve
\begin{equation}
\label{eq:xdot-case}
\dot x +\nu x = \frac{\partial H_0}{\partial \theta} \left(x,\frac{\theta}{K}\right) = \frac 12\frac{\sigma^2}{\sqrt{\gamma^2(x) - 2
      \sigma^2\theta/K}}.
\end{equation}
A solution requires $\dot x + \nu x > 0$, as should be expected from \eqref{eq:1}. We then obtain
\begin{equation}
  \label{eq:70}
  \vartheta(x,\dot x) = \frac{\theta}{K} = \frac{\gamma^2(x)}{2\sigma^2} 
  - \frac{\sigma^2}{8(\dot x + \nu x)^2} 
\end{equation}
and get
\begin{equation}
  \label{eq:71}
    \mathcal{L}(x,\dot x)  = \frac{K}{8\sigma^2} 
    \frac{|2\gamma(x) (\dot x + \nu x) - \sigma^2|^2}{\dot x + \nu x}
\end{equation}
whenever $\dot x + \nu x > 0$.

\begin{figure}
  \includegraphics[width=233pt]{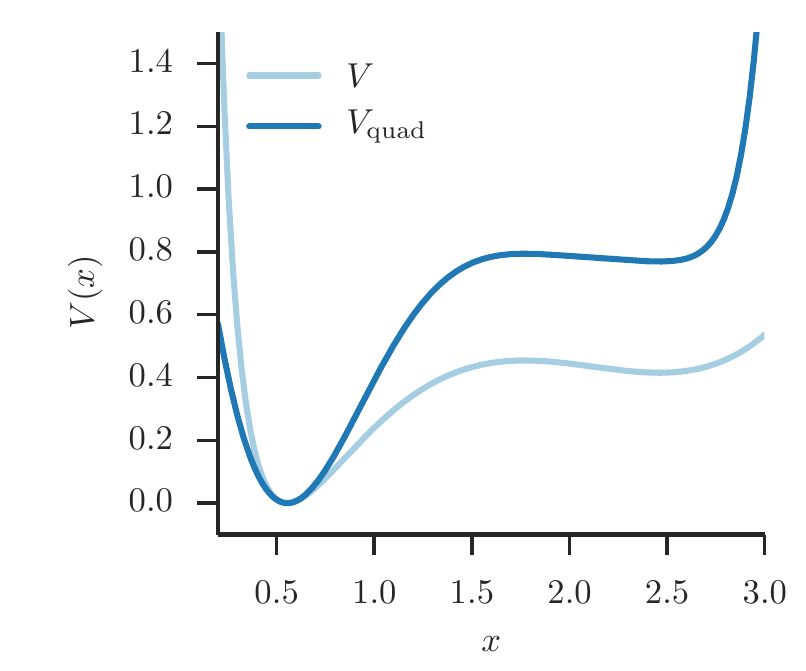}\caption{The potential
    $V(x)/K$ and the one obtained from a quadratic approximation of
    the Hamiltonian $V_{\text{quad}}(x)/K$, see \eqref{eq:60q}, for
    $\gamma (x) = x^4/10 - x^2 +3$, $\nu=1$, $\sigma = \sqrt{3}$. The
    quadratic potential obtained from the quadratic approximation is
    quite different from the actual potential.\label{fig:potential}}
\end{figure}

\begin{figure}
  \includegraphics[width=233pt]{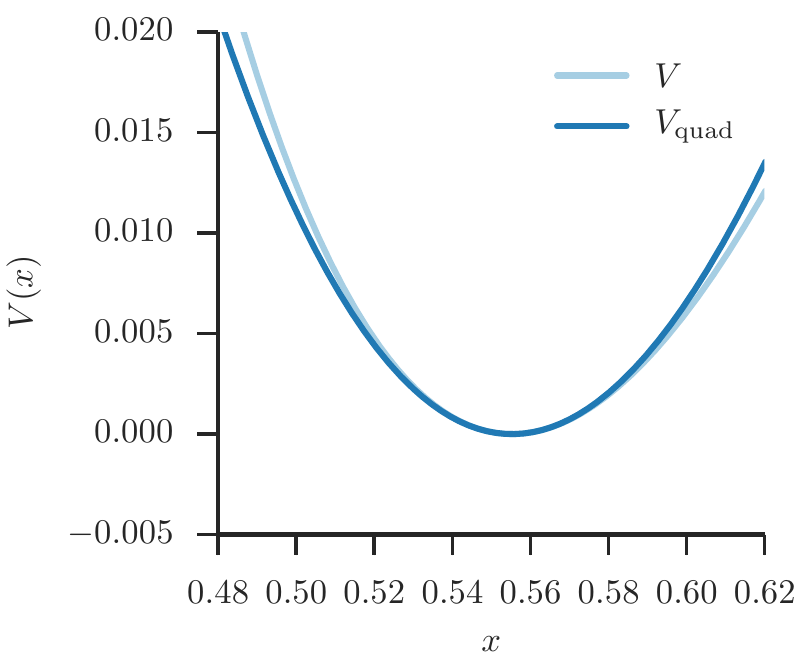}\caption{The
    potential $V(x)/K$ and the one obtained from a quadratic
    approximation of the Hamiltonian $V_{\text{quad}}(x)/K$ in the
    vicinity of the main attractor.\label{fig:potential_zoom}}
\end{figure}

The quasi-potential is given by \eqref{eq:91}, using~\eqref{eq:H-Yk-independent} and \eqref{eq:H_0-case} it is easy to see that this implies that either
$\partial_x V \equiv V' =0$, or
\begin{equation}
  \label{eq:60}
  \frac{V'(x)}{K} = \frac{\nu  x \gamma(x) - \tfrac12\sigma^2 }{ \nu^2 x^2}
\end{equation}
This result should be compared with the one obtained from the
quadratic approximation to $H$,
\begin{equation}
  \label{eq:61}
  H_{\text{quad}} = -\nu x \theta +K\left[\frac{\sigma^2\theta/K}{2\gamma(x)} +
  \frac{\sigma^4 (\theta/K)^2}{4  \gamma^3(x)}\right],
\end{equation}
for which we deduce
\begin{equation}
  \label{eq:60q}
  \frac{V_{\text{quad}}'(x)}{K} = \frac{4 \gamma^3(x)}{\sigma^4} \left(\nu x- 
    \frac{\sigma^2}{ 2\gamma(x)}\right)
\end{equation}
The potentials $V(x)$ and $V_{\text{quad}}(x)$ are different in
general. To give a concrete example, consider the case
$\gamma (x) = x^4/10 - x^2 +3$, $\nu=1$, $\sigma = \sqrt{3}$, which
leads to bistability of the slow process $x(t)$. The potentials $V$
and $V_{\text{quad}}$ are represented in figure
\ref{fig:potential}. The extrema of $V_{\text{quad}}$ are also extrema
of $V$, and the second derivatives of these two potentials are the
same at these extrema, as should be expected.  Figure
\ref{fig:potential_zoom} illustrates this last point for the potential
minima. However, we see that the global shape of the potentials are
very different, and in particular the energy barrier between the
attractors of $V_{\text{quad}}$ is almost twice the one between the
attractors of $V$, which means that the probability of rare
transitions obtained from the quadratic approximation in this case
will be much lower than the actual one.


\subsection{The limit as $K\to\infty$}
\label{sec:Kinf}

Interestingly, we can corroborate the results obtained in
Sec.~\ref{sec:alp0} by taking the limit as $K\to\infty$ first.  If we
define
\begin{equation}
  \label{eq:2}
  E = \frac1K \sum_{k=1}^K Y_k^2 
\end{equation}
it is easy to see that \eqref{eq:1} can be rewritten as
\begin{equation}
  \label{eq:3}
   \begin{cases}
    \dot X = E - \nu X\\
    \displaystyle dE  = - \frac{2}{\alpha}\gamma(X) E dt +
    \frac{1}{\alpha} \sigma^2 dt +
    2\frac{\sigma}{\sqrt{\alpha K}} 
    \sqrt{E} \, dW
  \end{cases}
\end{equation}
where we used the identity
\begin{equation}
  \label{eq:4}
  \frac{1}{K} \sum_{k=1}^K Y_k dW_k = K^{-1/2}
    \sqrt{E} \, dW \qquad \text{(in law)}
\end{equation}
Since the noise term in~\eqref{eq:3} is small when $K$ is large, we
can use large deviation theory to analyze the behavior of the system in
the limit as $K\to\infty$. The Freidlin-Wentzell action associated
with~\eqref{eq:3} reads
\begin{equation}
  \label{eq:18}
  I_T^\alpha(E,x) = \frac{\alpha}{8\sigma^2}\int_0^T 
  \frac{|\dot E + 2\gamma(x)E/\alpha - \sigma^2/\alpha|^2}{E}dt
\end{equation}
if $E = \dot x+\nu x$ and $I_T(E,x)=\infty$ otherwise. Letting
$\alpha\to0$ and keeping only the leading order term gives
\begin{equation}
  \label{eq:72}
  I_T^\alpha(E,x) \sim
  \frac{1}{8\sigma^2\alpha}\int_0^T 
  \frac{|2\gamma(x) (\dot x + \nu x) - \sigma^2|^2}{\dot x +
    \nu x}dt
\end{equation}
which is consistent with~\eqref{eq:71} since they both imply that the
probability weight on paths is roughly 
\begin{equation}
  \label{eq:5}
  \exp\left(- \frac{K}{8\sigma^2\alpha } \int_0^T 
    \frac{|2\gamma(x) (\dot x + \nu x) - \sigma^2|^2}{\dot x + \nu x} dt\right) 
\end{equation}
when $\alpha$ is small and/or $K$ is large. 

\subsection{Two-dimensional generalization}
\label{sec:2D}

\begin{figure}
  \begin{center}
    \includegraphics[width=233pt]{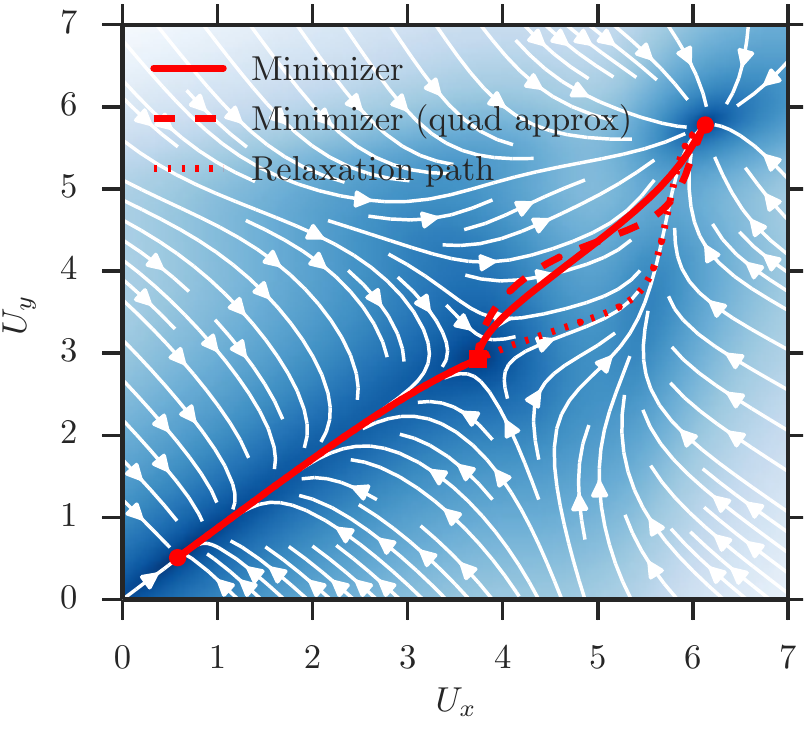}
  \end{center}
  \caption{Dynamics of the coupled slow-fast system ODE model for
    $\kappa=0.2$. The arrows denote the direction of the deterministic
    flow, the color its magnitude. The red solid line depicts the
    minimizer of the actual action associated with the Hamiltonian
    in~\eqref{eq:H-slowfast2d}, the red dashed line the minimizer of
    the quadratic approximation of this Hamiltonian, and the red
    dotted line the relaxation paths from the saddle via the limiting
    equation~\eqref{eq:LLN-fastslow2d}. Red markers are located at the
    fixed points (circle: stable; square:
    saddle).\label{fig:slowfast2d}}
\end{figure}

To illustrate the impact that the non-quadratic nature of the
Hamiltonian has on the pathway of the transition, let us now consider
the following generalization of~\eqref{eq:1}:
\begin{equation}
  \label{eq:slowfast2d}
  \begin{cases}
    \begin{aligned}
      \dot X_i &= -\beta_i X_i + \kappa {\textstyle\sum_{j=1}^2 }D_{ij} X_j +
      Y_i^2, \quad &i=1,2\\
      dY_i &= -\frac1\alpha \gamma(X_i) Y_i dt +
      \frac1{\sqrt{\alpha}}\sigma dW_i, \quad &i=1,2
    \end{aligned}
  \end{cases}
\end{equation}
with
\begin{equation}
  \label{eq:D}
  D_{11}=D_{22} = -1, \qquad D_{12}=D_{21} = 1
\end{equation}
The LLN equations for the system \eqref{eq:slowfast2d} are given by
\begin{equation}
  \label{eq:LLN-fastslow2d}
  \Dot{\bar X}_i= \frac{\sigma^2}{2\gamma(\bar X_i)} - \beta_i \bar X_i 
  + \kappa {\textstyle\sum_{j=1}^2 }D_{ij} X_j, \quad i=1,2.
\end{equation}
For the specific choice $\gamma(x) = (x-5)^2+1$, and $\beta_1=0.6$,
$\beta_2=0.3$ and $\sigma=\sqrt{10}$, the flow field associated
with~\eqref{eq:LLN-fastslow2d} is shown in figure
\ref{fig:slowfast2d}: it has the two stable fixed points (shown as red
circles in the figure) with one unstable critical point (shown as a
red square) in between.

\begin{figure}
  \begin{center}
    \includegraphics[width=233pt]{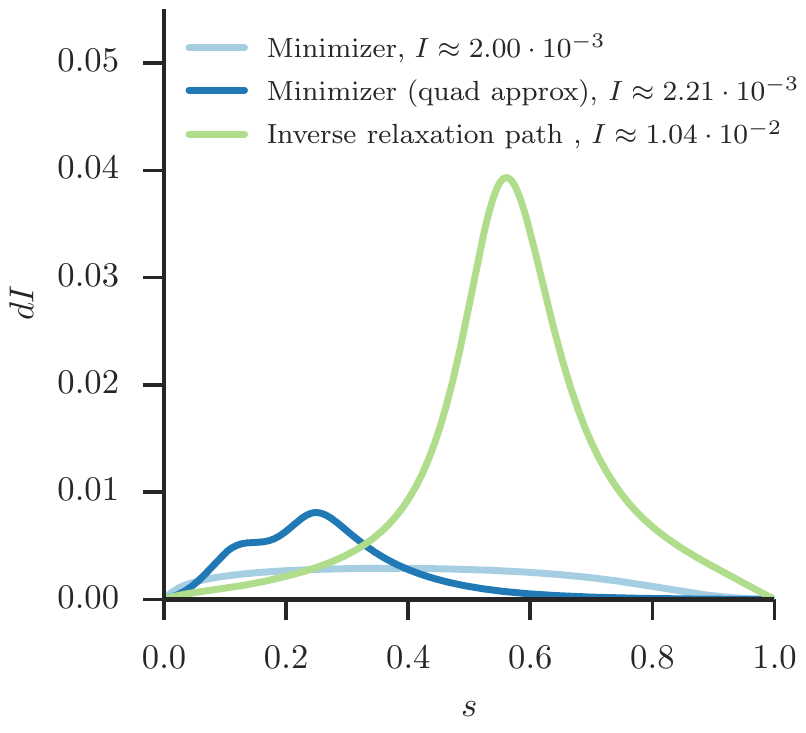}
  \end{center}
  \caption{Action density $dI$ (variation of the action per unit
    length) versus path length $s$, for the coupled slow-fast system
    ODE model in~(\ref{eq:slowfast2d}) for $\kappa=0.2$, for paths up
    to the saddle point. The action density is computed with respect
    to the full Hamiltonian \eqref{eq:H-slowfast2d} for the three
    trajectories depicted in figure \ref{fig:slowfast2d}, i.e. the
    minimizer of the actual action, the minimizer of the quadratic
    approximation of this action, and the relaxation pathway from the
    limiting equation~\eqref{eq:LLN-fastslow2d}.
  \label{fig:slowfast2daction}}
\end{figure}
The Hamiltonian associated with \eqref{eq:slowfast2d} can be written as
\begin{equation}
  \label{eq:H-slowfast2d}
  H(x, \theta) = \sum_{i=1}^2 h(x_i, \theta_i) - \sum_{i=1}^2 \beta_i
  x_i \theta_i  + \kappa \sum_{i,j=1}^2 \theta_j D_{ij} x_j
\end{equation}
with
\begin{equation}
  h(x, \theta) = \tfrac12 \left(\gamma(x) 
    - \sqrt{\gamma^2(x) - 2\sigma^2 \theta}\right)\,.
\end{equation}
A numerical computation of the transition trajectories between the two
stable fixed points was performed using the geometric minimum action
method GMAM~\cite{heymann2008geometric,heymann2008pathways} (building
on the method introduced in~\cite{e2004minimum}) and is shown in
figure \ref{fig:slowfast2d}: The red solid line depicts the minimizing
trajectory for the full Hamiltonian \eqref{eq:H-slowfast2d}, while the
dashed line represents the minimizing trajectory for a quadratic
approximation of $H(x,\theta)$, with clear differences between the
two. The respective probability of the minimizers can be seen by
evaluating the action along the trajectories, as shown in
Fig.~\ref{fig:slowfast2daction}: The action evaluated along its
minimizer is lower than evaluated along the minimizer of the quadratic
action. Also showed in dashed lines are the relaxation paths from the
unstable to the stable equilibrium points from the limiting
equation~\eqref{eq:LLN-fastslow2d}: These paths are followed by the
minimizing trajectories on the way down from the unstable critical
points (as they should: no noise is necessary for this part of the
transition paths) but not on the way uphill.

\section{Concluding Remarks}
\label{sec:homog}

We have investigated how large deviations affect the slow variables in
fast-slow systems via analysis of the large deviation principle (LDP)
that characterize their likelihood and pathways. For a specific class
of systems, we derived an algebraic equation for the Hamiltonian
involved in this LDP, and we discussed several situations in which
this equation can be solved explicitly. These results show that the
way rare events or infrequent transitions arise in fast-slow systems is
intrinsically different from the way they would arise if the dynamics
of the slow variables was approximated by an SDE -- these difference
stem from the fact that the Hamiltonian is non-quadratic in the
momenta in general. The examples treated in the present paper were
simple enough to allow for analytic treatment. However, we believe that
our results will be useful in more complicated situations, in which the
algebraic equation for the Hamiltonian will have to be solved
numerically.

\section*{Acknowledgments} 
E. V.-E. thank David Kelly for interesting discussions. The research leading to these results has received funding from the European Research Council under the European Union's seventh Framework Programme (FP7/2007-2013 Grant Agreement no. 616811) (F. Bouchet, and T. Tangarife).

\appendix

\section{ Derivation of the limiting equation~\eqref{eq:25} from the
  LLN}
\label{sec:llnderiv}

Here we derive the limiting equation~\eqref{eq:25} of the LLN by
formally taking the limit as $\alpha\to0$ on the backward Kolmogorov
equation~\eqref{eq:28}. To this end expand $u$ as
$u=u_0+\alpha u_1 +O(\alpha^2)$, insert this ansatz in~\eqref{eq:28},
and collect term of increasing power in $\alpha$. This gives the
hierarchy
\begin{equation}
  \label{eq:30}
  \begin{aligned}
    L_1 u_0 & = 0\\
    L_1 u_1 & = \partial_t u_0 - L_0 u_0\\
    & \vdots
  \end{aligned}
\end{equation}
The first implies that $u_0$ is a only a function of $x$ and not of $y$, or equivalently 
\begin{equation}
  \label{eq:31}
  Pu_0 = u_0.
\end{equation}
Since $L_1$ is
not invertible ($PL_1 = 0$), the second equation requires a
solvability condition, which reads
\begin{equation}
  \label{eq:32}
  0 = \partial_t Pu_0 - P L_0 u_0 = \partial_t u_0 - P L_0P u_0 
\end{equation}
It is easy to see that $PL_0P = F(x) \cdot \partial_x$,
i.e. \eqref{eq:32} is the backward Kolmogorov equation of the limiting
ODE~\eqref{eq:25}. 

\section{ Derivation of the CLT equation~\eqref{eq:33}}
\label{sec:cltderiv}

To derive the linear SDE~\eqref{eq:33} of the CLT, notice that,
using~\eqref{eq:25}, \eqref{eq:19} can be rewritten as
\begin{equation}
  \label{eq:34}
  \begin{cases}
    \displaystyle \Dot{\tilde \xi} = \frac{1}{\sqrt{\alpha}}\tilde
    f(\bar X, Y) +
    \partial_x f(\bar X,Y) \tilde \xi +
    O(\sqrt{\alpha})\\[8pt]
    \displaystyle dY = \frac1\alpha b(\bar X,Y) dt +
    \frac1{\sqrt{\alpha}} \partial_x b(\bar X, Y) \tilde \xi +
    \frac1{\sqrt{\alpha}} \sigma(\bar X, Y) dW(t) + O(1)
  \end{cases}
\end{equation}
This means that the joint process $(\bar X,\tilde \xi, Y)$ is Markov
with generator $L' = L_0'+ \alpha^{-1/2} L_{\frac12} + \alpha^{-1} L_1 +
O(\alpha^{3/2})$ where $L_1$ is defined in~\eqref{eq:26} and
\begin{equation}
  \label{eq:35}
  \begin{aligned}
    &L_0' =  F(\bar x) \cdot \partial_{\Bar x} + \partial_x f(\bar
    x,y) \xi \cdot \partial_{\xi} + \text{operator in $y$}\\
    &L_{\frac12} = \tilde f(\bar x,y)\cdot \partial_\xi+ \partial_x b(\bar
    x, y) \xi \cdot \partial_y
  \end{aligned}
\end{equation}
Letting
\begin{equation}
  \label{eq:29b}
  v(t,\bar x,\xi,y) = \EE^{\bar x,\xi,y} g(\bar X(t),\tilde \xi(t))
\end{equation}
this function satisfies the backward Kolmogorov equation
\begin{equation}
  \label{eq:28b}
  \partial_t v = L'_0 v + \frac1{\sqrt{\alpha}} L_{\frac12} v +
  \frac1\alpha L_1 v + \text{higher order terms}, \qquad v(0) = g
\end{equation}
Formally expand $v$ as $v=v_0 +\sqrt{\alpha} v_{\frac12}+\alpha v_1
+O(\alpha^{3/2})$, insert this ansatz in~\eqref{eq:28b}, and collect
term of increasing power in $\alpha$:
\begin{equation}
  \label{eq:30b}
  \begin{aligned}
    L_1 v_0 & = 0\\
    L_1 v_{\frac12} & = - L_{\frac12} v_0\\
    L_1 v_1 & = \partial_t v_0 - L'_0 v_0 - L_{\frac12} v_{\frac12}\\
    \cdots
  \end{aligned}
\end{equation}
The first equation implies that $v_0 = P v_0$, i.e. $v_0$ is a
function of $\bar x$ and $\xi$ only. The solvability condition for the
second equation is automatically satisfied since $P\tilde f = 0$
implies that $P L_{\frac12} P =0$.  Therefore, the solution to this
equation is
\begin{equation}
  \label{eq:37}
  v_{\frac12} = - L_1^{-1} L_{\frac12}P v_0
\end{equation}
where $L_1^{-1}$ denotes the pseudo-inverse of $L_1$. Alternatively,
this solution can also be expressed as
\begin{equation}
  \label{eq:38}
  v_{\frac12} = \int_0^\infty d\tau\, e^{\tau L_1}   L_{\frac12} Pv_0
\end{equation}
Using this expression in the solvability condition for the third
equation in~\eqref{eq:30b} finally gives the evolution equation for
$v_0$:
\begin{equation}
  \label{eq:39}
  \partial_t v_0 = PL'_0P v_0 + P L_{\frac12}\int_0^\infty d\tau\, e^{\tau
    L_1}
  L_{\frac12} P v_0
\end{equation}
The first term at the right hand side is explicitly
\begin{equation}
  \label{eq:40}
  PL'_0Pv_0  = \bar F(\bar x) \cdot \partial_{\bar x} v_0 + 
  B_1(\bar x) \xi \cdot \partial_\xi v_0
\end{equation}
where $B_1(x)$ is
\begin{equation}
  \label{eq:41B1}
    B_1(x) = \int_{\RR^n} \partial_x f(x,y) \mu_x(dy)
\end{equation}
The second term at the right hand side of~\eqref{eq:39} is
\begin{equation}
  \label{eq:42}
  P L_{\frac12}\int_0^\infty d\tau\, e^{\tau L_1} L_{\frac12} P v_0
  = B_2(\bar x) \xi \cdot \partial_\xi v_0 +
  A(\bar x) : \partial_\xi \partial_\xi v_0
\end{equation}
where $A(x)$ is the matrix defined in~\eqref{eq:43} and 
\begin{equation}
  \label{eq:41B2}
    B_2(x)  = \int_0^\infty d\tau \int_{\RR^n} \left(\partial_y
    \EE^y f(x,\tilde Y^x(\tau))\right) \partial_x b(x,y) \mu_x(dy) 
\end{equation}
Inserting~\eqref{eq:40} and~\eqref{eq:42} in \eqref{eq:39} shows that
this equation is indeed the backward Kolmogorov equation of the joint
process governed by~\eqref{eq:25} and~\eqref{eq:33}.

\bibliographystyle{plain}
\bibliography{newbib2}

\end{document}